%% file: main.tex
\documentclass[conference]{IEEEtran}
\IEEEoverridecommandlockouts
\usepackage[T1]{fontenc}
\usepackage{hyperref}
\usepackage[anythingbreaks]{breakurl}
\usepackage{algorithm}
\usepackage{algorithmicx}
\usepackage{algpseudocode}
\usepackage{graphicx}
\usepackage{textcomp}
\usepackage{amsthm,amsfonts,amssymb}
\usepackage{mathtools}
\usepackage{multirow}
\usepackage{comment}
\usepackage{lettrine}
\usepackage{url}
\usepackage{fancybox}
\usepackage{multicol}
\usepackage{booktabs}
\usepackage{threeparttable}
\usepackage{makecell}

\usepackage{xcolor}
\usepackage{colortbl} 
\definecolor{cambridgeblue}{rgb}{0.64, 0.76, 0.68}

\usepackage{subcaption}

\usepackage{pifont} 

\definecolor{darkgreen}{RGB}{0, 100, 0}
\usepackage{enumitem}

\usepackage[justification=centering]{caption}\usepackage{cleveref}

\algblockdefx{Event}{EndEvent}[1]%
{\textbf{upon event }#1 \textbf{do }}%
{\textbf{}}

\theoremstyle{definition} 

\newtheorem{defi}{Definition}
\theoremstyle{definition}

\newtheorem*{theorem*}{Theorem}

\algblockdefx{Function}{EndFunction}[1]%
{\textbf{Function }#1 }%
{\textbf{}}
\newcommand\algvspace{\vspace{0.6\baselineskip}}
\newcommand{\ccomment}[1]{\textcolor{gray}{#1}}



\begin{document}

\title{\textsc{Juno}: Aggregated Vector Consensus for Optimal Asynchronous Common Subset}

\author{\IEEEauthorblockN{Liangrong Zhao}
\IEEEauthorblockA{\textit{Monash University} \\
liangrong.zhao1@monash.edu}
\and
\IEEEauthorblockN{Qin Wang}
\IEEEauthorblockA{\textit{CSIRO Data61, Australia} \\
qin.wang@data61.csiro.au}
\and
\IEEEauthorblockN{Joseph K. Liu}
\IEEEauthorblockA{\textit{Monash University} \\
joseph.liu@monash.edu}
\and
\IEEEauthorblockN{Jiangshan Yu$^{*}$}
\IEEEauthorblockA{\textit{University of Sydney} \\
jiangshan.yu@sydney.edu.au}
\IEEEcompsocitemizethanks{
\IEEEcompsocthanksitem $^{*}$Corresponding author
}
}

\maketitle

\begin{abstract}

In this paper, we propose \textit{aggregated vector consensus}, a new vector consensus primitive designed for asynchronous networks. The primitive achieves agreement by outputting a vector of values aggregated from independent process inputs.
We then introduce \textsc{Juno}, an asynchronous common subset (ACS) protocol that fully implements our aggregated vector consensus to attain optimal $\mathcal{O}(n^2)$ message complexity. 

We further implement and evaluate \textsc{Juno} in comparison with the legacy HoneyBadgerBFT and the state-of-the-art Dory. Experiment results demonstrate its efficacy and efficiency. Our protocol demonstrates an average throughput performance improvement of 93\% compared with HoneyBadgerBFT and a 47\% improvement compared with Dory.
Notably, our study makes significant progress in addressing the gap in applying vector consensus protocol in fully asynchronous networks.


\end{abstract}

\smallskip



\input{paper/introduction}
\input{paper/preliminaries}

\input{paper/design}

\input{paper/protocol}

\input{paper/simulation}

\input{paper/literature}

\section{Conclusion}

In this paper, we introduce a novel primitive called \textit{aggregated vector consensus} (AVC) for constructing efficient and optimal asynchronous common subset protocols. AVC allows for the collection of multiple independent instance inputs and their aggregation into a single vector-based output.

Building on AVC, we design and implement \textsc{Juno}, a practical ACS protocol. \textsc{Juno} achieves optimal $O(n^2)$ message complexity. Experimental evaluations demonstrate that it outperforms HoneyBadgerBFT (by 93\%) and Dory (47\%).

\bibliographystyle{unsrt}
\bibliography{main}




\end{document}

%% file: paper/introduction.tex
\section{Introduction}\label{sec-intro}

Byzantine fault tolerance (BFT) protocols~\cite{castro1999practical} allow a distributed network of \textit{n} processes (i.e., other consensus nodes) to reach consensus despite the presence of \textit{f} malicious (i.e., Byzantine) processes.
Traditional BFT protocols enable the correct processes to agree on a single value. However, this approach is insufficient for modern distributed systems with multiple inputs and outputs. A leader might propose a set of \( n \) transactions (e.g., in the context of blockchain~\cite{duan2022recent}), but correct processes may only want to accept a subset of these transactions rather than accepting or rejecting the entire block.

The vector consensus problem generalizes this concept by allowing processes to agree on a vector of values rather than a single value~\cite{doudou1998muteness}. In vector consensus, processes propose their values and converge on a single vector that should include the majority of the values proposed by the correct processes. 
However, vector consensus was originally explored in synchronous networks or relied on additional mechanisms like muteness detectors or ``wormholes''~\cite{doudou1998muteness}\cite{neves2005solving}. Most vector consensus primitives, in the worst cases, depend on partial synchrony~\cite{neves2005solving}\cite{cachin2020anonymity}\cite{correia2006consensus}, which can be problematic when applied directly to asynchronous networks. Processes in asynchronous networks might have different responses (or votes from an agreement perspective) due to arbitrary delays. If more than \( f \) processes disagree on a transaction in the proposed set, the consensus process might be aborted or fail, compromising liveness. To our knowledge, there have been very limited efforts to apply vector consensus in asynchronous networks~\cite{moniz2008ritas}.

The other closely related concept is asynchronous common subset (ACS), which requires that
the output of each correct process contains \( n-f \) values, with at least \( n-2f \) of them originally proposed by correct processes.  
ACS is a fundamental primitive for constructing asynchronous BFT protocols~\cite{miller2016honey}\cite{ben1994asynchronous}\cite{cachin2001secure}\cite{guo2020dumbo}\cite{miller2016honey}\cite{duan2018beat}\cite{gao2022dumbo}\cite{zhang2022pace}. An ACS protocol typically consists of two phases: the \textit{broadcast} phase, where processes disseminate transactions to others, and the \textit{agreement} phase, where a common subset is decided by collecting a sufficient number of input values from different processes.

Most ACS constructions adhere to a few fixed design patterns and primarily rely on two paradigms (see details in Table~\ref{tab:pattern}). Ben-Or, Kelmer, and Rabin~\cite{ben1994asynchronous} introduced the practical ACS framework known as the \textit{BKR} paradigm by incorporating multiple reliable broadcast (RBC) and asynchronous binary agreement (ABA) instances. This framework has inspired subsequent studies, including HoneyBadgerBFT~\cite{miller2016honey}, BEAT~\cite{duan2018beat}, PACE~\cite{zhang2022pace} and EPIC~\cite{liu2020epic}. Cachin, Kursawe, Petzold, and Shoup~\cite{cachin2001secure} proposed the \textit{CKPS} paradigm, which shifted the construction paradigm by leveraging asynchronous multivalued validated Byzantine agreement (MVBA). This approach was followed by a series of works on the Dumbo family of protocols~\cite{guo2020dumbo}\cite{gao2022dumbo}\cite{lu2020dumbo}.

MVBA-based constructions provide optimal time complexity, partially addressing BKR's performance bottleneck due to its linear message complexity (per transaction) and constant time complexity. Compared to BKR, which requires \(n\) parallel ABA invocations, MVBA can utilize vectors with multiple values as inputs, enabling a single agreement execution to select a vector of values rather than a single value~\cite{lu2020dumbo}. This leads to an asymptotic improvement in message complexity~\cite{guo2022speeding}.

Nonetheless, similar to vector consensus, MVBA typically assumes that one of the input vectors will be selected as the final decision. In other words, the consensus protocol aims to unanimously decide which of the proposed vectors to accept. Such a design overlooks the fact that correct processes in an asynchronous network might have different votes due to unbounded delays. This creates a challenging scenario where votes from a correct process may not be entirely accurate. Consequently, a single input vector may not yield the best consensus outcome because even vectors from correct processes can contain incorrect values due to poor network conditions, particularly in asynchronous settings.

Compared with MVBA-based ACS primitives with vector features, the independent relationships among transactions are more effectively featured and utilized in BKR with \(n\) separate ABA instances. PACE enhances this approach by replacing the traditional ABA with a re-proposable ABA, which biases the agreement preference towards one, considering that correct processes might initially have ``wrong'' votes~\cite{zhang2022pace}. This motivates us to explore the BKR paradigm. However, BKR constructions have been criticized for their redundant message complexity due to the \textit{n} parallel RBC and ABA instances, resulting in an overall message complexity of \(\mathcal{O}(n^3)\). This is asymptotically higher than the \(\mathcal{O}(n^2)\) complexity in MVBA.

\smallskip
\noindent\textbf{What is aggregated vector consensus?} Inspired by the parallel ABA design in the BKR paradigm, we propose a new aggregated vector consensus (AVC) primitive where the final agreement outcome is aggregated rather than selected from input vectors. Processes can individually vote and reach a consensus for each value in the final vector consensus outcome. Aggregated vector consensus is an agreement where a process $p_i$ (where $1 \leq i \leq n$) inputs a vector $V_i$ and correct processes attempt to decide on a vector $V$ while satisfying related properties (formally defined in Definition~\ref{def-aggreVector}). In particular, the agreed $V$ could be either a proposed $V_i$, or a vector of aggregated elements from different proposed vectors.

\underline{\textit{Road to use AVC to achieve a better ACS.}}


We aim to develop an ACS protocol that enhances normal-case performance through independent value processing, while also maintaining the optimal message complexity achieved by recent MVBA-based ACS protocols~\cite{zhang2022pace}\cite{guo2022speeding}\cite{duan2023fin}.
We start from reviewing the technical roots of constructing ACS protocols and try to adopt the vector design to lower the overall complexity. We show our efforts in both two phases (namely, broadcast and agreement).

Replacing the existing RBC with a lighter provable broadcast (PB), as inspired by previous works~\cite{abraham2019asymptotically}\cite{guo2022speeding}, is challenging in the BKR paradigm due to the absence of all-to-all communication in PB, which is essential for achieving the \textit{agreement} property required in the subsequent agreement phase. Specifically, RBC guarantees that all correct processes can eventually deliver the same value if there is a value to deliver (i.e., \textit{agreement}), while PB only ensures that if two correct processes deliver, they deliver the same value (i.e., \textit{consistency}). This ``property loss'' presents challenges for ACS constructions, as the subsequent agreement phase typically relies on the agreement property to ensure that all correct processes have the same delivery outcomes.

MVBA protocols, such as VABA~\cite{abraham2019asymptotically} and Speeding Dumbo~\cite{guo2022speeding}, require either additional PB invocations, more communication rounds, or extra phases like the recovery phase~\cite{guo2022speeding}. We aim to design an agreement protocol that allows us to replace RBC with PB without adding any extra communication overhead.


We shift our focus to the agreement phase and discover that it is feasible to achieve a reduced overall message complexity of $\mathcal{O}(n^2)$ by employing aggregated vectors in consensus procedures. In traditional ABA settings, each process participates in all $n$ ABA instances. At each step of the agreement phase, every one of the $n$ processes sends $n$ vote messages to each of the $n$ recipients, resulting in a message complexity of $\mathcal{O}(n^3)$.

In contrast, we adopt an AVC implementation where, at each step of the agreement phase, each of the $n$ processes sends only one vote message to each of the $n$ recipients, reducing the message complexity to $\mathcal{O}(n^2)$. This is accomplished by aggregating a process's $n$ vote messages — each containing a vote intended for a single ABA instance — into a single message that includes a vector of $n$ binary votes for all $n$ ABA instances. Each element of the vector represents an individual vote for a specific ABA instance.

\smallskip
\noindent\textbf{Contributions.} Based on this design principle, we formally deliver a series of research outputs.

\vspace{0.15em}
\underline{\textit{A new primitive: aggregated vector consensus.}}
\vspace{0.15em}

We introduce a new concept called the \textit{aggregated vector consensus} (AVC) primitive (Sec.\ref{sec-avc}). In AVC, the final agreement is formed by aggregating input vectors instead of selecting from them. Each process votes on individual values within these vectors to reach a consensus on the final vector. 


Our study of AVC fills the gap by (i) applying a vector consensus protocol in fully asynchronous networks; and (ii) exploring the potential of adapting the PB module to the BKR paradigm (Table~\ref{tab:pattern}). Surprisingly, our approach simplifies the complex communication patterns of parallel ABA designs, reducing the required communication steps compared to state-of-the-art vector-based MVBA implementations.

\input{paper/tab-compr}

\vspace{0.15em}
\underline{\textit{A new ACS construction riding on the new primitive.}}
\vspace{0.15em}

We propose \textsc{Juno}, an instantiated construction that fully implements our aggregated vector consensus primitive (Sec.\ref{sec-design}). \textsc{Juno} consists of two refined protocols.

Firstly, we design an agreement protocol based on our newly proposed aggregated vector consensus. Our protocol adopts the vector input approach to have a single agreement instance, encapsulating the \textit{n} parallel agreement instances. The overall message complexity of the agreement phase matches the optimal $\mathcal{O}(n^2)$ for \textit{n} instances. Our aggregated vector consensus can merge \(n\) messages into a single message, reducing communication costs for \(n-1\) authenticators for each process in each round.

Secondly, we present a \textit{provable broadcast (PB) protocol} tailored for our aggregated vector agreement. We design a lightweight broadcast protocol with linear message complexity to match the agreement protocol. Our PB protocol facilitates the following agreement phase, where the delivery statuses recorded during the broadcast determine the agreement inputs.

\vspace{0.15em}
\underline{\textit{Full implementation of \textsc{Juno} with evaluations.}}
\vspace{0.15em}

We developed, evaluated, and examined our protocol \textsc{Juno}, addressing aspects of performance, complexity, and security (Sec.\ref{sec-eva}). In our implementation, \textsc{Juno} demonstrates significant improvements in throughput and latency. We conducted a comparative simulation, assessing \textsc{Juno} against prominent protocols such as the open-source HoneyBadgerBFT (improved by \textbf{93\%}) and Dory (by \textbf{47\%}). \textsc{Juno} is capable of tolerating up to \textit{f} Byzantine faults in a system comprising ($3f+1$) total processes. The message complexity remains \(\mathcal{O}(n)\) per transaction and \(\mathcal{O}(n^2)\) when processing \(n\) parallel transactions. 

%% file: paper/tab-compr.tex
 \begin{table}[!hbt]

   \centering
   \caption{Comparisons for ACS \& Asynchronous BFTs}
   \label{tab:pattern}
  \renewcommand{\arraystretch}{1.30}
   \footnotesize
   \setlength{\tabcolsep}{4pt} 
   \begin{threeparttable}
   \resizebox{\linewidth}{!}{
   \begin{tabular}{c|c |c c | c  }
     \toprule
    \multicolumn{1}{c}{\textbf{Protocols}} & \multicolumn{1}{c}{\textbf{Paradigm}} & \multicolumn{1}{c}{\textbf{Broadcast}} & \multicolumn{1}{c}{\textbf{Agreement}} & {\textbf{Message}} \\ 
     \midrule
     {HoneybadgerBFT}~\cite{miller2016honey} &  \cellcolor{green!15}  BKR & RBC & ABA & \(\mathcal{O}(n^3)\)   \\ 
     BEAT~\cite{liu2020epic} &  \cellcolor{green!15}  BKR & RBC & ABA  &    \(\mathcal{O}(n^3)\) \\
     EPIC~\cite{liu2020epic} &  \cellcolor{green!15}  BKR & RBC & ABA &    \(\mathcal{O}(n^3)\) \\
     {PACE}~\cite{zhang2022pace} &  \cellcolor{green!15}  PACE\tnote{*} & RBC & rABA\tnote{**} & \(\mathcal{O}(n^3)\)   \\ 

     \midrule
     
     {VABA}~\cite{abraham2019asymptotically} & CKPS &  \cellcolor{green!15}  PB & MVBA & \(\mathcal{O}(n^2)\) \\ 
     {Dumbo}~\cite{guo2020dumbo} & CKPS & RBC & MVBA & \(\mathcal{O}(n^3)\)   \\
     {Speeding Dumbo}~\cite{gao2022dumbo} & CKPS &  \cellcolor{green!15}  PB & MVBA & \(\mathcal{O}(n^2)\)  \\
     {Dory}~\cite{zhang2022dory} & CKPS &  \cellcolor{green!20}  PB & MVBA & \(\mathcal{O}(n^2)\)  \\ 
     \midrule
     {\textbf{\textsc{Juno}}} & \cellcolor{green!15} BKR &  \cellcolor{green!15}  \cellcolor{green!15}  PB &   AVC & \(\mathcal{O}(n^2)\)  \\ 
     \midrule   
     \multicolumn{2}{c|}{} & \multicolumn{2}{c|}{\textbf{Design Pattern}} & \multicolumn{1}{c}{\textbf{Complexity}} \\
   \end{tabular}
   }
     \begin{tablenotes}
     \item[*] The paradigm for Pace is similar to BKR. 
     \item[**] rABA stands for the reproposable ABA.
   \end{tablenotes}
   \end{threeparttable}

 \end{table}

%% file: paper/preliminaries.tex
\section{ACS and Aggregated Vector Consensus}
\label{sec-avc}

We revisit BKR-based ACS protocols and introduce a new primitive, denoted as the \textit{aggregated vector consensus} (AVC).

\smallskip
\subsection{Asynchronous Common Subset} 

In ACS protocols, each process broadcasts a value, typically known as a transaction in the context of blockchains, as input. All correct processes will output a common subset $\mathbb{S}$ of these inputs, ensuring that transactions from at least $n-f$ processes are included. Formally, a secure ACS protocol satisfies the following properties~\cite{miller2016honey}:
 
\begin{itemize}
\item \textbf{ACS-validity}: If a correct node outputs $\mathbb{S}$,
then $|\mathbb{S}|$$\geq$\textit{n-f}, and $\mathbb{S}$ contains the inputs of at least \textit{n-2f} correct nodes.
\item \textbf{ACS-agreement}: If a correct node outputs $\mathbb{S}$, then every correct node outputs $\mathbb{S}$.
\item \textbf{ACS-totality}: If \textit{n-f} correct nodes receive an input, then all correct nodes produce an output.
\end{itemize}

In general, ACS protocols consist of two phases: (i) \textit{broadcast}: transactions are disseminated to all other processes; and (ii) \textit{agreement}: correct processes reach an agreement regarding which transactions should be included in the final subset. 


\subsection{The Broadcast Phase}
\label{subsec-broadcast}

\subsubsection{\underline{Reliable broadcast}}
Transactions have to be disseminated as the votes in the subsequent agreement phase rely on their delivery status. Most of the ACS protocols employ the well-explored RBC as their broadcast primitive for its reliability in asynchronous networks. Bracha's original double-echo RBC protocol results in a total communication overhead of $\mathcal{O}(n^2)$ as it requires rounds of all-to-all communication to satisfy the following properties~\cite{cachin2011introduction}. 

\begin{itemize}
\item \textbf{RBC-validity}: If a correct process \textit{p} broadcasts a message \textit{m}, then every correct process eventually delivers \textit{m}.
\item \textbf{RBC-none-duplication}: No correct process delivers message \textit{m} more than once. 
\item \textbf{RBC-integrity}: If a correct process delivers a message \textit{m} with a correct sender \textit{p}, then the message \textit{m} was previously broadcast by \textit{p}.
\item \textbf{RBC-agreement}: If some correct process delivers a message \textit{m}, then every correct process eventually delivers \textit{m}.
\end{itemize}

\smallskip
\subsubsection{\underline{Provable broadcast.}}
Recent works on MVBA-based ACS protocols have explored the option of using a much cheaper provable broadcast primitive to bring down the high communication cost of the broadcast phase. Compared with the all-to-all communication pattern in the reliable broadcast, the provable broadcast's all-to-one pattern only incurs a linear communication cost. Different sets of properties can be achieved by varying the number of sequential invocations of provable broadcast. In general, provable broadcast primitives can obtain the following properties~\footnote{While different papers assign varied names and definitions, we adhere to a consistent set of reliable broadcast properties for easy comparison in the following sections~\cite{cachin2011introduction}}:

\begin{itemize}
\item \textbf{PB-validity}: If a correct process \textit{p} broadcasts a message \textit{m}, then every correct process eventually delivers \textit{m}.
\item \textbf{PB-none-duplication}: No correct process delivers message \textit{m} more than once. 
\item \textbf{PB-integrity}: If a correct process delivers a message \textit{m} with a correct sender \textit{p}, then the message \textit{m} was previously broadcast by \textit{p}.
\item \textbf{PB-consistency}: If some correct process delivers a message \textit{m} and another correct process delivers a message \textit{m'}, then \textit{m} = \textit{m'}.
\end{itemize}

\subsection{The Agreement Phase}
\label{subsec-agreement}

A Byzantine agreement protocol enables processes to reach a mutual agreement in the presence of malicious parties. Typically, a Byzantine agreement invocation decides the acceptance of a single value proposed by some process. Equivalently, the protocol aims to select a correct value and ensure that all correct processes eventually accept this value as the sole outcome. As a result, multiple agreement invocations are necessary if different values proposed by different processes need to be accepted simultaneously.

\subsubsection{\underline{Vector consensus}}
The concept of using vectors in Byzantine consensus was first introduced in the vector consensus problem~\cite{doudou1998muteness} as a reduction method for atomic broadcast primitives. In the vector consensus problem, correct processes aim to agree on a set of values represented as a vector.

A similar design was also introduced in MVBA. It replaced the generalized multi-valued input with a vector containing multiple values, allowing the agreement protocol to output one of the selected vectors from inputs rather than a single value.
Formally, the vector consensus problem is defined as~\cite{doudou1998muteness}:

\begin{defi}[Vector consensus]\label{def-VectorConsensus} 
Vector consensus is a form of agreement where each process inputs a value \textit{v} and correct processes attempt to decide on a vector \textit{V} of values so that the following properties are satisfied: 
    \begin{itemize}
    \item \textbf{VC-agreement}: If two correct processes decide \textit{V} and \textit{V'}, then \textit{V} = \textit{V'}.
    \item \textbf{VC-termination}: All correct processes eventually decide. 
    \item \textbf{VC-validity}: If a correct process decides \textit{V}, then \textit{V} satisfies the following conditions:
        \begin{itemize}
        \item For every \(1 \leq i \leq n\), if process \(p_i\) is correct, then \(\textit{V}[i]\) is either the \(v_i\), initial value of \(p_i\), or \(\perp\).
        \item at least \textit{f+1} elements of the vector \textit{V} are the initial values of correct processes.
        \end{itemize}
    \end{itemize}
\end{defi}

Compared to ACS, vector consensus has a slightly different validity property: it requires the final vector output to contain at least \( f+1 \) correct values, whereas ACS only requires the inclusion of \( n-f \) correct values. The difference is caused by their ways of handling inputs and outputs: in vector consensus, each process inputs a single value and outputs a vector of values, while in MVBA, each process inputs a vector and selects one of the input vectors as the output. Another major difference is the definition of validity.


\smallskip
\subsubsection{\underline{Validity}}

Validity is a critical property for an asynchronous Byzantine agreement as it determines the criteria of correctness of the agreement outcome. There are two common validity definitions: \textit{strong validity} and \textit{weak validity}~\cite{cachin2001secure}.

\begin{itemize}
\item \textbf{Strong validity}: If a correct process decides \textit{v}, at least one correct process proposes \textit{v}.

\item \textbf{Weak validity}: If all correct processes propose \textit{v}, then every correct process that terminates decides \textit{v}.
\end{itemize}

Weak validity ensures a minimum termination condition: if all correct processes unanimously propose the same value, this value is guaranteed to be accepted. In contrast, strong validity requires a stronger liveness condition, considering cases where correct processes might propose different values. It ensures the correctness of the agreement decision as long as it is originally proposed by a correct process.  

MVBA further assumes an \textit{external validity} to ensure the protocol's output is not only consistent and reliable but also appropriate and meaningful within the context of the application using it~\cite{cachin2001secure}.

\begin{itemize}
\item \textbf{External validity}: If a correct process decides \textit{v}, \textit{v} satisfies the external predict function.
\end{itemize}

However, when adapting to a vector agreement, existing validity definitions overlooked a key point: the elements within a vector are typically considered independent. Each element in the final decision vector should ideally be determined individually, rather than by selecting one of the input vectors as the agreement outcome (e.g., in Dumbo-MVBA \cite{lu2020dumbo}).

In vector consensus, the validity of each value in the final decision vector follows the principle of \textit{strong validity}. A value is deemed valid only if it is the input from a correct process. However, the validity definition in vector consensus presents challenges when directly applied in an ACS protocol, where each element in a vector typically represents a vote on the inclusion of a specific transaction. Correct processes might cast \textit{incorrect} votes due to varying delivery statuses, caused by unpredictable delays in an asynchronous network.

\subsection{New Primitive: Aggregated Vector Consensus}
\label{subsec-agreement}

We accordingly propose a new set of property definitions that better align with the requirements of the ACS protocol and the features of vector consensus.

\begin{defi}[Aggregated Vector Consensus]\label{def-aggreVector} 
Aggregated vector consensus is an agreement where a process \(p_i\), \(1 \leq i \leq n\), inputs a vector \(V_i\) and correct processes attempt to decide on a vector \textit{V} so that the following properties are satisfied: 

\begin{itemize}
\item \textbf{AVC-agreement}: If two correct processes decide \textit{V} and \textit{V'}, then, \textit{V} = \textit{V'}.
\item \textbf{AVC-termination}: All correct processes will eventually decide.
\item \textbf{AVC-validity}: If a correct process decides \(\textit{V}\), then, for each \( 1 \leq i \leq n \), there exists a correct process \( p_j \) such that \(\textit{V}[i] = V_j[i]\).
\end{itemize}

\end{defi}

Our definition of validity ensures that the overall agreement vector's validity relies on the validity of each individual element of proposed vectors, rather than just considering the vector as a whole. Moreover, our aggregated vector consensus allows correct processes to propose conflicting values, as long as the final decided value has been proposed by at least one correct process and will be unanimously agreed upon. 

Intuitively, our definition can be viewed as a generalized version of vector consensus, where each process proposes \textit{n} inputs instead of just one. If process \(p_i\)'s input vector \(V_i\) contains only a single value \(V_i[i]\) with all other elements in \(V_i\) being \(\perp\), our definition aligns closely with the classic definition of vector consensus given in Definition~\ref{def-VectorConsensus}.


%% file: paper/design.tex
\section{\textsc{Juno} in a Nutshell}
\label{sec-design}

\subsection{General System Model}\label{subsec-model}

\noindent \textbf{Processes.}
 We consider a network $\Pi$ = \{$P_0, P_1,..., P_{n-1}$\} of \textit{n} processes with identifiers known to all in the network. We assume that processes are authenticated, i.e., they can rely on a digital signature scheme to authenticate each other's messages.
 We consider up to \textit{f}\textit{$\leq$ $\lfloor\frac{n-1}{3}\rfloor$} processes may be Byzantine. 
 Byzantine processes may behave arbitrarily, but cryptographic primitives remain secure. 

 \vspace{0.3em}
 \noindent \textbf{Asynchronous network.}
 We consider an asynchronous network, where messages sent by correct processes are eventually delivered after an unknown delay. Processes are connected in a distributed manner where each process is able to communicate with any other process via a direct connection. 
 
 The communication link between each pair of processes ensures that no message from the correct sender is lost, duplicated or indefinitely delayed. However, messages can be arbitrarily delayed or reordered.

\begin{figure*}[htp]
    \centering
    \includegraphics[width=0.9\textwidth]{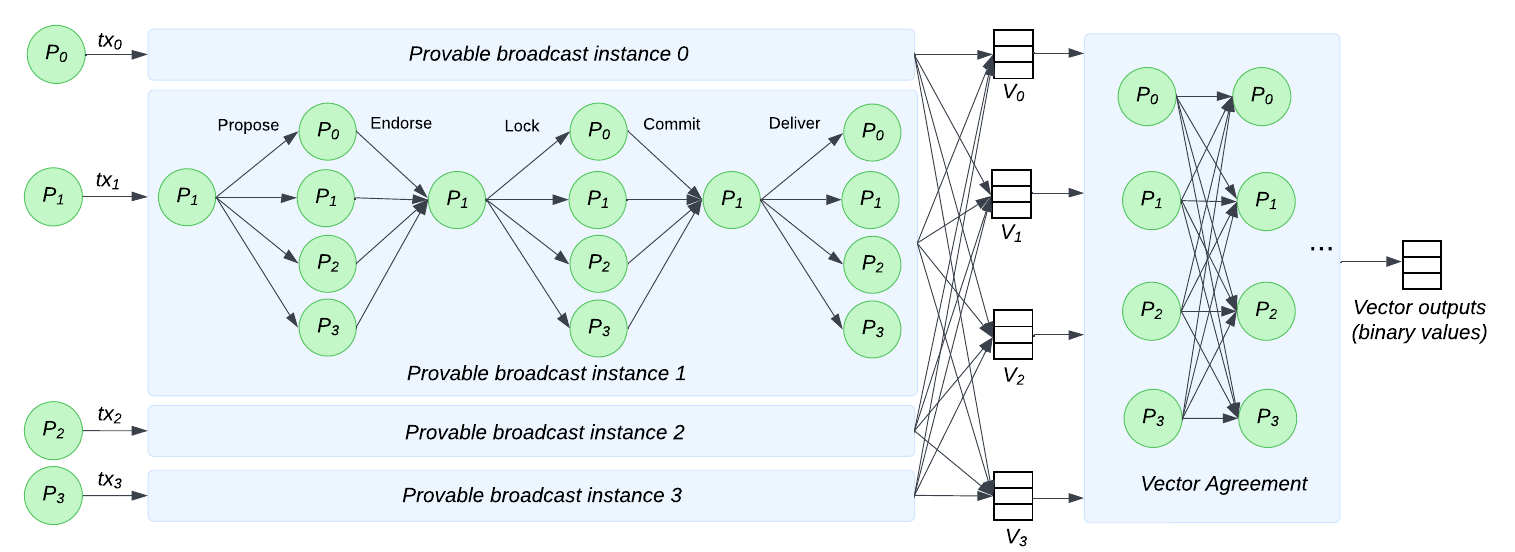} 
    \caption{Overall structure of \textsc{Juno}.}
    \label{fig:overall}
\end{figure*}

\subsection{\textsc{Juno} Overview}
We propose a new ACS protocol, \textsc{Juno}, which incorporates parallel provable broadcast instances along with a vector agreement instance (the structure diagram given in Fig.\ref{fig:overall}) where each process inputs a transaction and outputs the same vector. Our protocol introduces two new features in its design.

\vspace{0.3em}
\underline{\textit{Design-\ding{172}: Obtaining a linear broadcast instance.}} 
\vspace{0.3em}

There are \textit{n} parallel broadcast instances, with each process serving as the sender (referred to as the \textit{primary} from a consensus view) of an instance. In each instance, the primary initiates a transaction by broadcasting a \textcolor{violet}{$\mathsf{Propose}$} message. Processes receiving messages respond with \textcolor{violet}{$\mathsf{Endorse}$} after verification. Once a threshold of \textsc{Endorse} messages is reached, the primary broadcasts a \textcolor{violet}{$\mathsf{Lock}$} message. A process receiving a valid \textcolor{violet}{$\mathsf{Lock}$} message will return a \textcolor{violet}{$\mathsf{Commit}$} message. Similarly, the primary broadcasts a \textcolor{violet}{$\mathsf{Deliver}$} message after collecting enough \textcolor{violet}{$\mathsf{Commit}$} messages, and processes receiving a valid \textcolor{violet}{$\mathsf{Deliver}$} message consider the initial transaction as delivered. This design ensures linear message complexity per transaction.

\vspace{0.3em}
\underline{\textit{Design-\ding{173}: Reducing message complexity for agreement.}}
\vspace{0.3em}

Departing from the conventional BKR paradigm that runs a separate ABA instance for each transaction, \textsc{Juno} utilizes the proposed aggregated vector consensus design to combine a process's \textit{n} votes for the parallel ABA instances into a single vector, thereby reducing the message complexity per transaction from \(\mathcal{O}(n^2)\) to \(\mathcal{O}(n)\).


\subsection{(Adjusted) Provable Broadcast}

Prior to commencing binary agreement (BA) for binary decisions, transactions have to be disseminated as the votes in the agreement phase rely on their delivery statuses. We propose to replace the RBC with a more cost-effective PB protocol, which demonstrates linear complexity in only four steps (one-all-one-all-one).
This replacement eliminates RBC's all-to-all communication phases, resulting in a reduction in message complexity.

However, it hasn't been without trade-offs. The lack of all-to-all communication makes the non-primary processes in PB constrained to receiving messages solely from the primary. Consequently, in the event of a faulty primary, not all properties of reliable broadcast can be guaranteed in the PB, even with hardware assistance. Notably, the achievement of \textit{RBC-agreement} becomes unattainable, as there is no assurance that all correct processes will eventually deliver a message when the primary's correctness is not guaranteed~\cite{cachin2011introduction}. 

To complement it, our design bypasses the need for the \textit{RBC-agreement} property not provided by PB and only relies on its \textbf{PB-consistency}.
Compared with the \textit{RBC-agreement} property, the consistency property does not guarantee \textit{totality}, which ensures that if a message is delivered to a correct process, every correct process eventually receives the same message~\cite{cachin2011introduction}. There might be a situation where one correct process delivers a message while another correct process has not yet received it from the sender. 

VABA's solution is to introduce a \textit{key-lock-commit} mechanism that can be implemented with four consecutive PB executions. This is to ensure all correct processes in a later view's broadcast phase can \textit{convince} the left-out processes to proceed to the latest view after they lock themselves for a view change after ``commit''~\cite{abraham2019asymptotically}. 
Speeding Dumbo manages to reduce two invocations by introducing two additional steps of all-to-all communication in the agreement phase. They also introduce an effective ``recovery'' phase to allow processes that do not receive a message to retrieve a copy of this message from others using a ``help me'' function.

To address this property loss, we propose a provable broadcast primitive to deliver a transaction to its \textit{best capability}. If a correct process is unable to deliver a transaction in a timely manner (presumably due to a faulty sender), it will cast a vote based on the stage it has completed with the transaction and let the subsequent agreement protocol decide on it.
In our design, we keep a simple message exchange pattern in the broadcast phase while ensuring that such reduction does not impact and can facilitate the progression of the agreement protocol. Our provable broadcast has the following stages:

\begin{itemize}
    \item \textbf{Propose}. The sender broadcasts a transaction.
    \item \textbf{Endorse}. A process returns an endorsement message to the sender if the transaction is valid.
    \item \textbf{Lock}. The sender broadcasts a lock message if it has received \textit{2f+1} endorsements.
    \item \textbf{Commit}. A process returns a commit message to the sender if the lock message is valid.
    \item \textbf{Deliver}. The sender broadcasts a deliver message if it has received \textit{2f+1} commit messages.
\end{itemize}



\subsection{Aggregated Vector Agreement}
Recall the voting rules in the traditional BKR paradigm: a process votes 1 in an ABA instance only after the corresponding transaction is valid and delivered via RBC. A process votes 0 if the transaction fails to be delivered, and there are \textit{n-f} transactions that have been delivered. 
Similarly, the states for a transaction, determined by its delivery status in the broadcast instance, also play a crucial role in facilitating our agreement protocol. The state of a transaction determines the process's votes for that transaction. 

\begin{figure}[htp]
    \centering
    \includegraphics[width=0.4\textwidth]{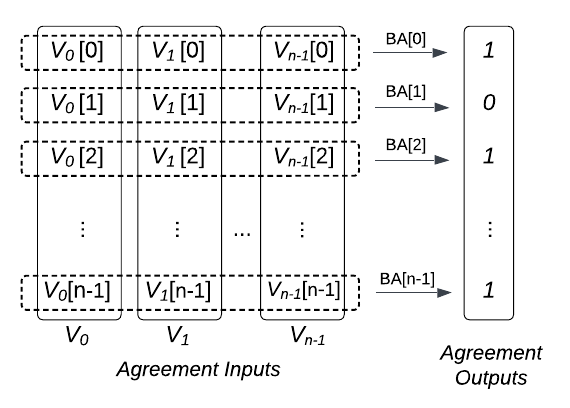}
    \vspace*{-0.1cm}
    \caption{Our agreement protocol with aggregated vector consensus (i.e., vector inputs and outputs).} 
    \label{fig:vector}
    \vspace{-0.2cm}
\end{figure}

In contrast to the traditional ABA construction in ACS where each process casts a vote for every transaction, we merge the \textit{n} vote messages into a single message with a \textit{n}-sized vote vector, each element representing a vote for an individual Byzantine agreement for one of the \textit{n} transactions.
As shown in Fig~\ref{fig:overall}, the vector input allows \textsc{Juno} to merge the \textit{n} parallel agreement instances into one vector agreement instance where a process exchanges its \textit{n}-size vectors in a single message instead of sending \textit{n} messages to \textit{n} agreement instances.
Specifically, a binary decision is independently achieved for each index ranging from 0 to \textit{n-1} of the vectors, corresponding to \textit{n} ABA instances in the traditional BKR paradigm. As a result, the final output from our agreement protocol is a decision vector where each element represents the binary decision corresponding to a specific transaction. The binary decision for a transaction \textit{i} is derived from the \textit{i}-th element across all input vectors to the agreement protocol. As in Fig.\ref{fig:vector}: \textit{V}$_0$, \textit{V}$_1$,..., \textit{V}${_{n-1}}$ are vote vectors from processes \textit{P}$_0$, \textit{P}$_1$,..., and \textit{P}${_{n-1}}$, respectively. 

%% file: paper/protocol.tex
\section{\textsc{Juno} Construction}
\label{sec:Juno}

In this section, we present the full \textsc{Juno} protocol\footnote{We exclude the well-known~\cite{mostefaoui2014signature} procedures of termination and message processing in the pseudocode for simplicity.} given as pseudocode in Algorithm~\ref{algorithm:tpbft}.
\input{paper/codes}

\subsection{Provable Broadcast Protocol (\textbf{Step} \ding{172}-\ding{176})}

Our proposed PB protocol is structured into five distinct phases: \textit{propose}, \textit{endorse}, \textit{lock}, \textit{commit}, and \textit{deliver}. Each phase aligns with the type of broadcast message during its respective phase. Here, we outline the abstracted workflow of these steps.

\vspace{0.2em}
\noindent\textbf{Step-\ding{172}: Propose.} At each process \textit{P}$_i$, \textit{P}$_i$ generates a \textcolor{violet}{$\mathsf{Propose}$} message with its transaction attached and broadcasts it to all other processes.
    
\vspace{0.3em}
\noindent\textbf{Step-\ding{173}: Endorse.} Upon receiving the \textcolor{violet}{$\mathsf{Propose}$} message from \textit{P}$_i$, processes verify the signature. If valid, they generate an \textcolor{violet}{$\mathsf{Endorse}$} message and send it back to \textit{P}$_i$.
    
\vspace{0.3em}
\noindent\textbf{Step-\ding{174}: Lock.} Upon receiving \textit{2f+1} \textcolor{violet}{$\mathsf{Endorse}$} messages, \textit{P}$_i$ verifies the validity of these messages and creates a \textcolor{violet}{$\mathsf{Lock}$} message to broadcast.

\vspace{0.3em}
\noindent\textbf{Step-\ding{175}: Commit.} Upon receiving the \textcolor{violet}{$\mathsf{Lock}$} message from \textit{P}$_i$, processes verify the signature. If valid, they generate an \textcolor{violet}{$\mathsf{Commit}$} message and send it back to \textit{P}$_i$. 

\vspace{0.3em}
\noindent\textbf{Step-\ding{176}: Deliver.} Upon receiving \textit{2f+1} \textcolor{violet}{$\mathsf{Commit}$} messages, \textit{P}$_i$ verifies validity and creates a \textcolor{violet}{$\mathsf{Deliver}$} message to broadcast.

\begin{figure}[!htp]
    \centering
    \includegraphics[width=1.0\columnwidth]{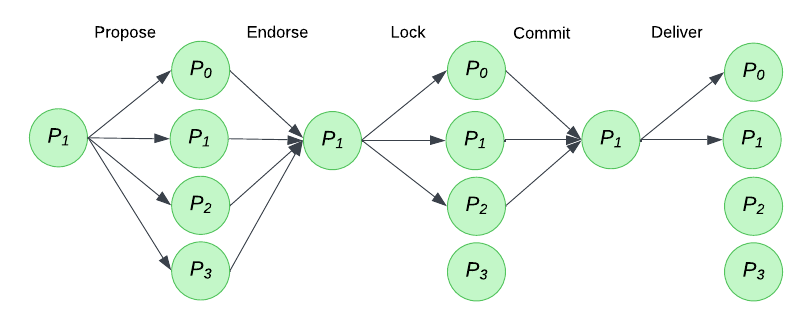}
    
    \caption{An example of the proposed provable broadcast.} 
    \label{fig:bcastExample}
    \vspace{-0.2cm}
\end{figure}

\subsection{Agreement Protocol with AVC (\textbf{Step} \ding{177}-\ding{179})}
\label{subsec:agreement}

An agreement protocol ensures that the correct processes share the same set of transactions for inclusion in ACS. Our agreement protocol is based on aggregated vector consensus. 

\vspace{0.2em}
\noindent \textbf{Step-\ding{177}: Generating a \textcolor{violet}{$\mathsf{Vote}$} message.}  At each process \textit{P}$_i$, \textit{P}$_i$ participates in all \textit{n} transaction broadcast instances and locally records the delivery statuses of these transactions. Among all \textit{n} broadcast instances, at least \textit{n-f} of them eventually guarantee successful delivery, as they are led by correct processes. When \textit{P}$_i$ has received at least \textit{n-f} valid \textcolor{violet}{$\mathsf{Deliver}$} messages in Round 1 (the first round), it will enter the agreement phase and produce a \textcolor{violet}{$\mathsf{Vote}$} message to broadcast with an \textit{n}-sized vector containing \textit{P}$_i$'s votes for all \textit{n} transactions.

It is worth emphasizing that although the final agreement decision is a vector of binary values, the votes cast by processes in the agreement protocol are not necessarily binary, as they go beyond just 0 and 1. In other words, the input vote vectors to the agreement protocol might contain non-binary values, while the output vector (consensus outcome) only contains 0 and 1. Specifically, there are four possible votes: 0, nil, null, and 1.

\begin{itemize}
    \item  A process votes 1 if the transaction is delivered in a valid \textcolor{violet}{$\mathsf{Deliver}$} message; 

    \item  A process votes 0 if the transaction is invalid or has never been received; 

    \item  A process votes nil when the transaction is received in a \textcolor{violet}{$\mathsf{Propose}$} message from the primary and verified to be valid, but no subsequent \textcolor{violet}{$\mathsf{Lock}$} message is received from the primary;

    \item  A process votes null when receiving a valid \textcolor{violet}{$\mathsf{Lock}$} message with \textit{2f+1} endorsements.
\end{itemize}

Votes are determined by a transaction's delivery status.
Typically, a vote of 1 represents acceptance of a transaction upon successful delivery, while a vote of 0 indicates rejection when the transaction is invalid or not received. The votes nil and null reflect intermediate states during the provable broadcast, indicating that the transaction has been received but not yet delivered.
We use the example given in Fig.\ref{fig:bcastExample} to further explain different vote conditions: 

\begin{itemize}

\item \textit{P}$_1$~is the sender (presumably malicious). \textit{P}$_0$ receives a valid \textcolor{violet}{$\mathsf{Deliver}$} message and the transaction is delivered. \textit{P}$_0$ will vote 1.

\item \textit{P}$_2$ receives a valid \textcolor{violet}{$\mathsf{Lock}$} message and returns a \textcolor{violet}{$\mathsf{Commit}$} message to the sender but no subsequent \textcolor{violet}{$\mathsf{Deliver}$} message is received.  \textit{P}$_2$ will vote null.

\item \textit{P}$_3$ receives a valid \textcolor{violet}{$\mathsf{Propose}$} message and returns a \textcolor{violet}{$\mathsf{Endorse}$} message to the sender but no subsequent \textcolor{violet}{$\mathsf{Lock}$} message is received. \textit{P}$_3$ will vote nil.

\item (not indicated in the figure) If \textit{P}$_3$ has never received a valid \textcolor{violet}{$\mathsf{Propose}$} message, it will vote 0.

\end{itemize}

\vspace{0.2em}
\noindent\textbf{Step-\ding{178}: Exchanging and updating the \textcolor{violet}{$\mathsf{Vote}$} message.} 
After broadcasting the \textcolor{violet}{$\mathsf{Vote}$} message, \textit{P}$_i$ awaits the reception of \textit{n-f} \textcolor{violet}{$\mathsf{Vote}$} messages and adjusts the vote messages accordingly. Similar to the reproposable ABA \cite{zhang2022pace} with a \textit{bias towards 1}, the updating logic is designed to give preference to 1, increasing its likelihood of becoming the final decision.
If \textit{P}$_i$ receives \textit{f+1} valid \textcolor{violet}{$\mathsf{Vote}$} message proposing 1 for a transaction (an element of the vector is 1) and \textit{P}$_i$'s own vote is not 1, it will change its vote to 1.

\vspace{0.3em}
\noindent \textbf{Step-\ding{179}: Flipping the coin and converging to agreements.} After collecting \textit{n-f} \textcolor{violet}{$\mathsf{Vote}$} messages, a common coin is tossed, and \textit{P}$_i$ endeavors to achieve an agreement when conditions are met.
Notably, the common coin value is optimistically set to 1 in the first round (Round 1) before outputting either 0 or 1 with equal probability in subsequent rounds. 

For each index \textit{x}, ranging from 0 to \textit{n-1}, in an \textit{n}-size vector representing \textit{n} processes:
\begin{itemize}
\item Upon receiving \textit{n-f} \textcolor{violet}{$\mathsf{Vote}$} messages where elements \textit{x} are all 1, \textit{P}$_i$ decides 1 for \textit{x} if the coin value is 1. 
\item Upon receiving \textit{n-f} \textcolor{violet}{$\mathsf{Vote}$} messages where elements \textit{x} are all 0, \textit{P}$_i$ decides 0 for \textit{x} if the coin value is 0. 
\item Upon receiving \textit{n-f} \textcolor{violet}{$\mathsf{Vote}$} messages where elements \textit{x} are all null, \textit{P}$_i$ decides 1 for \textit{x} if the coin value is 1. 
\end{itemize}

\subsection{Special Cases}

\smallskip 
\subsubsection{\underline{Blockage situation}}  We must account for scenarios where correct processes have different delivery statuses for the same transaction. If we were to simply split their votes between 0 and 1 based on whether the transaction was delivered, treating nil and null as 0, some correct processes voting 1 would eventually decide on 1, while others voting 0 would be unable to decide if faulty processes choose to equivocate. The root cause is that correct processes are likely to vote differently due to varying delivery statuses, and there is no guarantee that these delivery statuses will eventually converge to the same state. As a result, the ``reproposable agreement'' idea proposed by PACE does not address this issue.

The inclusion of nil and null votes addresses the property loss caused by the provable broadcast. This prevents a deadlock scenario that can compromise the \textit{termination} property, indefinitely stalling the protocol (breaking system \textit{liveness}).

A blockage situation might happen when votes from correct processes are divided between null and nil. It happens when some correct processes have received the \textcolor{violet}{$\mathsf{Lock}$} message while others have not. However, none of them have received the \textcolor{violet}{$\mathsf{Deliver}$} message to be converted to vote 1. Consequently, the correct processes might become stuck and fail to reach a consensus because none of the decisive conditions for consensus can be met as long as the Byzantine processes choose not to send any message.

In this case, we assert that the relevant transaction is valid, as a nil or null vote can be produced only when the corresponding \textcolor{violet}{$\mathsf{Propose}$} message is authenticated. As a result, both 1 and 0 can be acceptable as long as correct processes eventually converge to the same decision. The general logic to solve the blockage is: (i) a null vote can convert 0 or nil to null and eventually decide 1 if the common coin is 1; (ii) a 0 vote can convert nil to 0 and eventually decide 0 if the common coin is 0. 
Specifically, we solve this issue by enabling the following conditions if there is no decision made after Round 5: 
(i) A 0 or nil vote can be converted to null upon receiving \textit{f+1} null votes if the common coin is 1;
(ii) A 0 or nil vote can be converted to 1 upon receiving \textit{2f+1} null votes if the common coin is 1;
(iii) A 0 or nil vote can be converted to null upon receiving \textit{2f+1} nil votes if the common coin is 1;
(iv) A nil vote can be converted to 0 upon receiving \textit{2f+1} nil and 0 votes if the common coin is 0.

In summary, the agreement procedure tends toward deciding 1 when the common coin is 1 or toward deciding 0 when the common coin is 0. The null votes and nil votes interchangeably ``flip'' until the common coin facilitates one of the two decisive conditions, leading to a universal decision.


\smallskip 
\subsubsection{\underline{Network partition}} \label{protocol:partition}
Given the assumption of asynchrony, potential challenges arise where a correct process may fail to receive messages from other processes promptly due to network delays, causing a lag in progress compared to others. To address this, similar to protocols such as Hybster, we implement an arranging window mechanism~\cite{behl2017hybrids}. The window is defined by the counter number of the last achieved ACS, serving as the lower watermark, while the upper bound is set slightly higher. Only transactions falling within this interval are considered valid.
Correct processes broadcast \textcolor{violet}{$\mathsf{ACS}$} messages once an ACS is attained. These messages contain the counter number (as transactions within an ACS share the same counter number), along with the indexes and hash values of transactions. A lagging process can catch up by gathering a quorum of \textcolor{violet}{$\mathsf{ACS}$} messages and retrieving all transactions from finalized ACSs from other processes upon request.


In an asynchronous network, messages might be delivered in a different order than their initial orders.  Hence, if a process receives a new message with a higher counter number from a primary while it is working on a lower-numbered transaction, it temporarily stores the new message until preceding transactions from the same primary are finalized. Importantly, a process will only store transactions with counter numbers within the ordering window. Transactions with counter numbers higher than the high watermark will be discarded, as correct processes will issue any counter number before the previous one is finalized and their transactions can hardly have far-advanced counter numbers. This prevents a faulty process from overwhelming the external storage of correct processes by sending multiple messages with far subsequent counter numbers. Nonetheless, the counter number in an \textcolor{violet}{$\mathsf{ACS}$} message may exceed the ordering window due to network partitions, risking the rejection of the latest transactions by lagging processes. In this case, an up-to-date \textcolor{violet}{$\mathsf{ACS}$} message can synchronize the slow process and update its ordering window.

%% file: paper/codes.tex
\begin{algorithm*}
\footnotesize    
\caption{\textsc{Juno} Construction} 
\label{algorithm:tpbft}
\vspace{-0.5cm}
\begin{multicols}{2} 
\begin{algorithmic}[1]

\State \underline{\textbf{Parameters:}}
\algvspace
\State \hspace*{2em}$n$: total number of processes. 
\State \hspace*{2em}$f < n/3$: maximum number of Byzantine processes.
\algvspace


\State \underline{\textbf{Variables:}} \\ 
\algvspace
\hspace*{2em}\textit{T}: the transaction a primary intends to propose\\
\hspace*{2em}\textit{p}: primary's ID, this is also its instance's ID\\
\hspace*{2em}$\sigma$: the signature for message authentication\\
\hspace*{2em}\textit{coin}: the binary common coin value\\
\hspace*{2em}\textit{v}[]: vote values (vector)\\
\hspace*{2em}\textit{r}: the round number, starting from 1
\algvspace

\State \underline{\textbf{As a primary \textit{p}}}
\algvspace
\State \textbf{upon event} \textit{p} proposes transaction \textit{T} \textbf{do}
    \State \hspace*{2em}\textsl{Broadcast} $\langle$\textcolor{violet}{$\mathsf{Propose}$}, \textit{T}, $\sigma_p\rangle$
\algvspace

\State \textbf{upon event} receiving \textcolor{violet}{$\mathsf{Endorse}$} messages from \textit{2f+1} processes \textbf{do}
    \State \hspace*{2em}\ccomment{//check if there is any message with invalid signature}
    \State \hspace*{2em}\textbf{if} $\exists$ $\langle$\textcolor{violet}{$\mathsf{Endorse}$}, \textit{h}, $\sigma_j\rangle$ \textbf{where} verify($\sigma_j$) == false
    \State \hspace*{4em}break
    \State \hspace*{2em}\ccomment{//check if there is any duplicate messages}
    \State \hspace*{2em}\textbf{if} $\exists$ $\langle$\textcolor{violet}{$\mathsf{Endorse}$}, \textit{h}, $\sigma_j\rangle$ and $\langle$\textcolor{violet}{$\mathsf{Endorse}$}, \textit{h}, $\sigma_m\rangle$ \textbf{where} $\sigma_j$ == $\sigma_m$
    \State \hspace*{4em}break
    \State \hspace*{2em}\ccomment{//check if any message corresponds to the incorrect Propose message}
    \State \hspace*{2em}\textbf{if} $\exists$ $\langle$\textcolor{violet}{$\mathsf{Endorse}$}, \textit{h}, $\sigma_j\rangle$ \textbf{where} \textit{h} != hash($\langle$\textcolor{violet}{$\mathsf{Propose}$}, \textit{T}, $\sigma\rangle$)
    \State \hspace*{4em}break
    \State \hspace*{2em}\textit{h} $\leftarrow$ hash($\langle$\textcolor{violet}{$\mathsf{Propose}$}, \textit{T}, $\sigma_p\rangle$)
    \State \hspace*{2em}\textsl{Broadcast} $\langle$\textcolor{violet}{$\mathsf{Lock}$}, \textit{h}, $\sigma_p\rangle$
\algvspace

\State \textbf{upon event} receiving \textcolor{violet}{$\mathsf{Commit}$} messages from \textit{2f+1} processes \textbf{do}
    \State \hspace*{2em}\ccomment{//check if any message corresponds to the incorrect Lock message}
    \State \hspace*{2em}\textbf{if} $\exists$ $\langle$\textcolor{violet}{$\mathsf{Commit}$}, \textit{h}, $\sigma_j\rangle$ \textbf{where} \textit{h} != hash($\langle$\textcolor{violet}{$\mathsf{Lock}$}, \textit{h}, $\sigma\rangle$)
    \State \hspace*{4em}break
    \State \hspace*{2em}\ccomment{//the message duplication and signature verification remain the same, omitted here for simplicity}
    \State \hspace*{2em}\textit{h} $\leftarrow$ hash($\langle$\textcolor{violet}{$\mathsf{Lock}$}, \textit{h}, $\sigma_p\rangle$)
    \State \hspace*{2em}\textsl{Broadcast} $\langle$\textcolor{violet}{$\mathsf{Deliver}$}, \textit{h}, $\sigma_p\rangle$
\algvspace

\State \underline{\textbf{Each process \textit{i}}} 
\algvspace
\State \textbf{upon event} receiving a \textcolor{violet}{$\mathsf{Propose}$} message from \textit{p} \textbf{do}
    \State \hspace*{2em}\ccomment{//check if there is any message with invalid signature}
    \State \hspace*{2em}\textbf{if} $\langle$\textcolor{violet}{$\mathsf{Propose}$}, \textit{T}, $\sigma_p\rangle$ \textbf{where} verify($\sigma_p$) == false
    \State \hspace*{4em}break
    \State \hspace*{2em}\textit{h} $\leftarrow$ hash($\langle$\textcolor{violet}{$\mathsf{Propose}$}, \textit{T}, $\sigma_p\rangle$)
    \State \hspace*{2em}\textsl{Send} $\langle$\textcolor{violet}{$\mathsf{Endorse}$}, \textit{h}, $\sigma_i\rangle$ to \textit{p}
\algvspace

\State \textbf{upon event} receiving a \textcolor{violet}{$\mathsf{Lock}$} message from \textit{p} \textbf{do}
    \State \hspace*{2em}\textbf{if} $\langle$\textcolor{violet}{$\mathsf{Lock}$}, \textit{h}, $\sigma_p\rangle$ \textbf{where} verify($\sigma_p$) == false
    \State \hspace*{4em}break
    \State \hspace*{2em}\textbf{if} $\langle$\textcolor{violet}{$\mathsf{Lock}$}, \textit{h}, $\sigma_p\rangle$ \textbf{where} \textit{h} != hash($\langle$\textcolor{violet}{$\mathsf{Propose}$}, \textit{T}, $\sigma_p\rangle$)
    \State \hspace*{4em}break
    \State \hspace*{2em}\textit{h} $\leftarrow$ hash($\langle$\textcolor{violet}{$\mathsf{Lock}$}, \textit{h}, $\sigma_p\rangle$)
    \State \hspace*{2em}\textsl{Send} $\langle$\textcolor{violet}{$\mathsf{Commit}$}, \textit{h}, $\sigma_i\rangle$ to \textit{p}
    \State \hspace*{2em}\textit{v}$_i$[\textit{p}] $\leftarrow$ \textcolor{darkgray}{null}
\algvspace

\State \textbf{upon event} receiving a \textcolor{violet}{$\mathsf{Deliver}$} message from \textit{p} \textbf{do}
    \State \hspace*{2em}\textbf{if} $\langle$\textcolor{violet}{$\mathsf{Deliver}$}, \textit{h}, $\sigma_p\rangle$ \textbf{where} verify($\sigma_p$) == false
    \State \hspace*{4em}break
    \State \hspace*{2em}\textbf{if} $\langle$\textcolor{violet}{$\mathsf{Deliver}$}, \textit{h}, $\sigma_p\rangle$ \textbf{where} \textit{h} != hash($\langle$\textcolor{violet}{$\mathsf{Lock}$}, \textit{h}, $\sigma_p\rangle$)
    \State \hspace*{4em}break
    \State \hspace*{2em}\textit{v}$_i$[\textit{p}] $\leftarrow$ \textcolor{darkgray}{1}
\algvspace

\State \textbf{upon event} \textit{i} has at least \textit{n-f} \textcolor{darkgray}{$\mathsf{1}$} in its vote vector \textit{v} in round 1 \textbf{do}
    \State \hspace*{2em}\textbf{for} index \textit{x} \textbf{in} [\textit{0...n-1}] where no \textcolor{violet}{$\mathsf{Propose}$} message is received from process \textit{x}
    \State \hspace*{4em}\textit{v}$_i$[\textit{x}] $\leftarrow$ \textcolor{darkgray}{0}   
    \State \hspace*{2em}\textsl{Broadcast} $\langle$\textcolor{violet}{$\mathsf{Vote}$}, \textit{v}, \textit{r}, $\sigma_i\rangle$
\algvspace

\State \textbf{upon event} receiving \textit{n-f} \textcolor{violet}{$\mathsf{Vote}$} messages \textbf{do}
    \State \hspace*{2em}\ccomment{//for the first round, the coin value is 1}
    \State \hspace*{2em}\textit{coin} $\leftarrow$ \textsl{CommonCoin}(\textit{n-f $\sigma$}, \textit{r})
    
    \State \hspace*{2em} \ccomment{//try to reach an agreement for each instance}
    \State \hspace*{2em}\textbf{for} index \textit{x} \textbf{in} [\textit{0...n-1}] 
    \State \hspace*{4em}\textbf{if} $\exists$ \textit{f+1} \textcolor{violet}{$\mathsf{Vote}$} messages \textbf{where} \textit{v}[\textit{x}] == \textcolor{darkgray}{1} \textbf{then}
    \State \hspace*{6em}\textit{v}$_i$[\textit{x}] $\leftarrow$ \textcolor{darkgray}{1}
    
    \State \hspace*{4em}\textbf{if} $\exists$ \textit{2f+1} \textcolor{violet}{$\mathsf{Vote}$} messages \textbf{where} \textit{v}[\textit{x}] == \textcolor{darkgray}{null} \textbf{and} \textit{coin} == 1 \textbf{then}
    \State \hspace*{6em}\textit{v}$_i$[\textit{x}] $\leftarrow$ \textcolor{darkgray}{1}
    
    \State \hspace*{4em}\textbf{if} $\exists$ \textit{f+1} \textcolor{violet}{$\mathsf{Vote}$} messages \textbf{where} \textit{v}[\textit{x}] == \textcolor{darkgray}{null} \textbf{and} \textit{v}$_i$[\textit{x}] $\neq$ \textcolor{darkgray}{1} \textbf{and} \textit{coin} == 1 \textbf{then}
    \State \hspace*{6em}\textit{v}$_i$[\textit{x}] $\leftarrow$ \textcolor{darkgray}{null}

    \State \hspace*{4em}\textbf{if} $\exists$ \textit{2f+1} \textcolor{violet}{$\mathsf{Vote}$} messages \textbf{where} \textit{v}[\textit{x}] == \textcolor{darkgray}{nil} \textbf{and} \textit{v}$_i$[\textit{x}] == \textcolor{darkgray}{0} or \textcolor{darkgray}{nil} \textbf{and} \textit{coin} == 1 \textbf{then}
    \State \hspace*{6em}\textit{v}$_i$[\textit{x}] $\leftarrow$ \textcolor{darkgray}{null}

    \State \hspace*{4em}\textbf{if} $\exists$ \textit{2f+1} \textcolor{violet}{$\mathsf{Vote}$} messages \textbf{where} \textit{v}[\textit{x}] == \textcolor{darkgray}{nil} or \textcolor{darkgray}{0} \textbf{and} \textit{v}$_i$[\textit{x}] == \textcolor{darkgray}{nil} \textbf{and} \textit{coin} == 0 \textbf{then}
    \State \hspace*{6em}\textit{v}$_i$[\textit{x}] $\leftarrow$ \textcolor{darkgray}{0}

    \State \hspace*{4em}\textbf{if} $\exists$ \textit{2f+1} \textcolor{violet}{$\mathsf{Vote}$} messages \textbf{where} \textit{v}[\textit{x}] == \textcolor{darkgray}{1} \textbf{and} \textit{coin} == 1 \textbf{then}
    \State \hspace*{6em}\textsl{Decide} 1 for \textit{instance}$_x$
    \State \hspace*{4em}\textbf{if} $\exists$ \textit{2f+1} \textcolor{violet}{$\mathsf{Vote}$} messages \textbf{where} \textit{v}[\textit{x}] == \textcolor{darkgray}{0} \textbf{and} \textit{coin} == 0 \textbf{then}
    \State \hspace*{6em}\textsl{Decide} 0 for \textit{instance}$_x$

    \State \hspace*{2em}\textbf{if} all instances have been decided
    \State \hspace*{4em}\ccomment{//indexes and hash values of included transactions}
    \State \hspace*{4em}\textsl{Broadcast} $\langle$\textcolor{violet}{$\mathsf{ACS}$}, \textit{index[]}, \textit{hash[]}, $\sigma_i\rangle$ \ccomment{//the ACS is achieved}
    \State \hspace*{2em}\textbf{else}
    \State \hspace*{4em}\ccomment{//rebroadcast the $\mathsf{Vote}$ message for next round}
    \State \hspace*{4em} \textit{r}++
    \State \hspace*{4em}\textsl{Broadcast} $\langle$\textcolor{violet}{$\mathsf{Vote}$}, \textit{v}, \textit{r}, $\sigma_i\rangle$
\algvspace

\end{algorithmic}
\end{multicols}
\vspace{-0.3cm}
\end{algorithm*}

%% file: paper/simulation.tex
\section{Analysis and Evaluation}\label{sec-eva}

\subsection{Complexity Analysis}
In this paper, the message complexity refers to the total number of messages generated by correct processes and the communication complexity refers to the total number of bits exchanged among correct processes~\cite{cachin2001secure}. We achieve lower message complexity while communication complexity is aligned to other asynchronous BFT protocols.

We analyze both the \textit{message complexity}, indicating the total number of messages exchanged among processes, as well as the \textit{communication complexity}, which is the total number of bits exchanged.
\textsc{Juno}' message complexity is $\mathcal{O}$(\textit{$n^2$}) and $\mathcal{O}$(\textit{$n$}) per transaction while the overall communication complexity is asymptotically $\mathcal{O}$(\textit{$5n^2$ + $|m|n^2\log n$}). The \textit{$5n^2$} is due to \textit{n} parallel instances of provable broadcast with five communication steps (2.5 rounds) and \textit{$|m|n^2\log n$} comes from the agreement phase where \textit{$|m|$} is the \textcolor{violet}{$\mathsf{Vote}$} message size and $\log n$ indicates the number of \textcolor{violet}{$\mathsf{Vote}$} message exchange rounds required for reaching agreement. A \textcolor{violet}{$\mathsf{Vote}$} message includes a \textit{2n}-bit vote vector along with metadata such as process identifiers and signatures, which far predominate over the vote vector in size when \textit{n} is within a reasonable range. 









\subsection{Evaluation}
We benchmark \textsc{Juno} against the legacy HoneybadgerBFT and state-of-the-art Dory for assessing the performance of an ACS protocol~\cite{HBBFT}\cite{zhang2022dory}.



 \vspace{0.3em}
\noindent\textbf{Configuration.} Building on the open-source ResilientDB framework~\cite{resilientDB}, we implement both \textsc{Juno} and HoneyBadgerBFT and deploy them on Google Cloud. We also deploy the open source implementation\footnote{\url{https://github.com/xygdys/Dory-BFT-Consensus}} of Dory~\cite{zhang2022dory} on Google Cloud as another baseline. We deploy the protocol with a geo-distributed network configuration comprising \textit{E2-standard-8} machines with 8 vCPUs and 32GB memory, deployed across several regions: Sydney, Hong Kong, and Atlanta. The measured experimental settings of the network are presented in Table~\ref{tab:rrt}. The transaction size is set to 250 bytes, and each data point represents the average results over 20 runs.



\begin{table}[!hbt]
\caption{Round trip time \& bandwidth between each region}
\small
\label{tab:rrt}
\renewcommand{\arraystretch}{1.2} 
\setlength{\tabcolsep}{3pt} 
\resizebox{\linewidth}{!}{
\begin{tabular}{@{}lccc@{}}
\toprule
                     & \textbf{Sydney-HK} & \textbf{HK-Atlanta} & \textbf{Sydney-Atlanta} \\ \midrule
\textbf{Round trip} (ms) & 164          & 229  & 121            \\
\textbf{Bandwidth} (Mbps)     & 57.3         & 137.4         & 59.7           \\ \bottomrule
\end{tabular}
}
    \begin{tablenotes}
      \scriptsize
      \item[] \textbf{Abbrev.} HK for Hong Kong

     \end{tablenotes}
\end{table}

\begin{figure}[!ht]
\centering
    \begin{subfigure}[b]{0.49\textwidth}
        \centering
        \includegraphics[width=\linewidth]{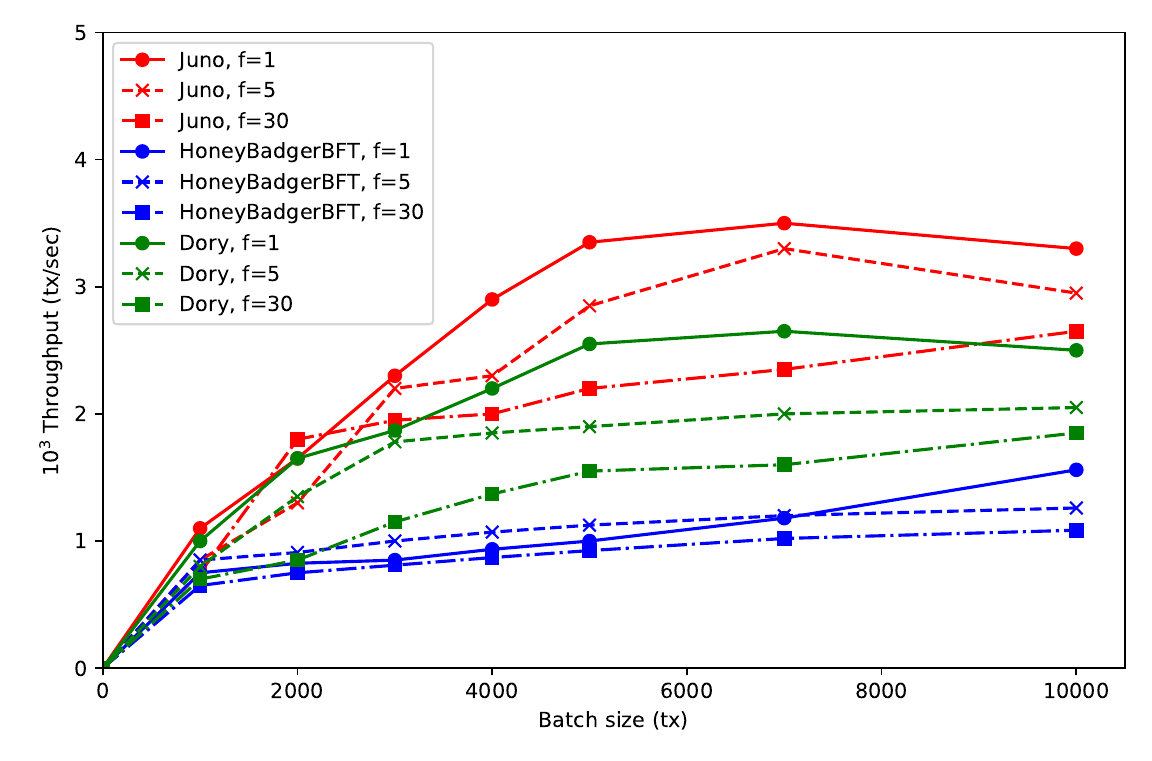}
        \caption{Throughput for different batch sizes}
        \label{fig:tput_geo}
    \end{subfigure}
    \hfill
    \vspace{0.1em}
    \begin{subfigure}[b]{0.49\textwidth}
        \centering
        \includegraphics[width=\linewidth]{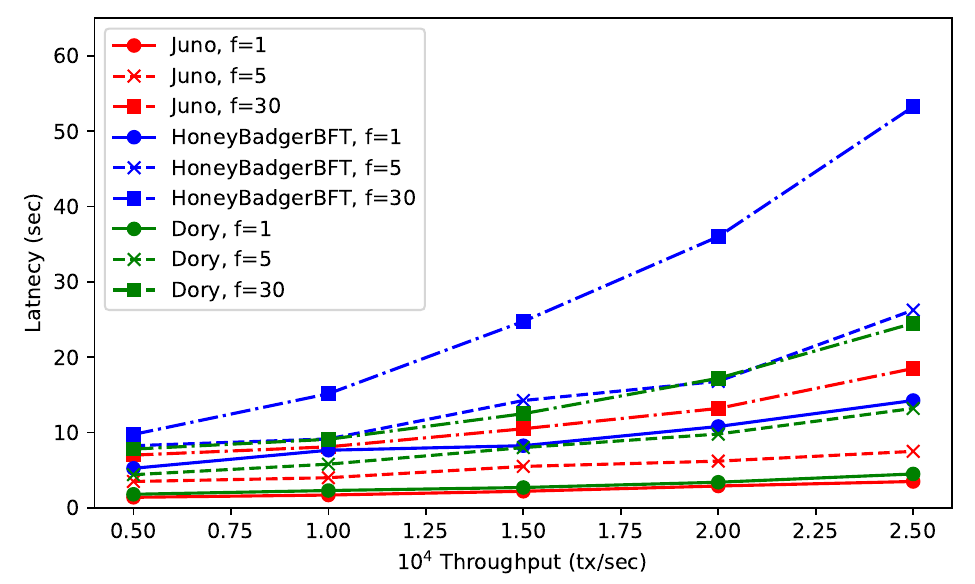}
        \caption{Latency vs throughput}
        \label{fig:latency}
    \end{subfigure}
    \hfill
    \begin{subfigure}[b]{0.53\textwidth}
        \centering
        \includegraphics[width=\linewidth]{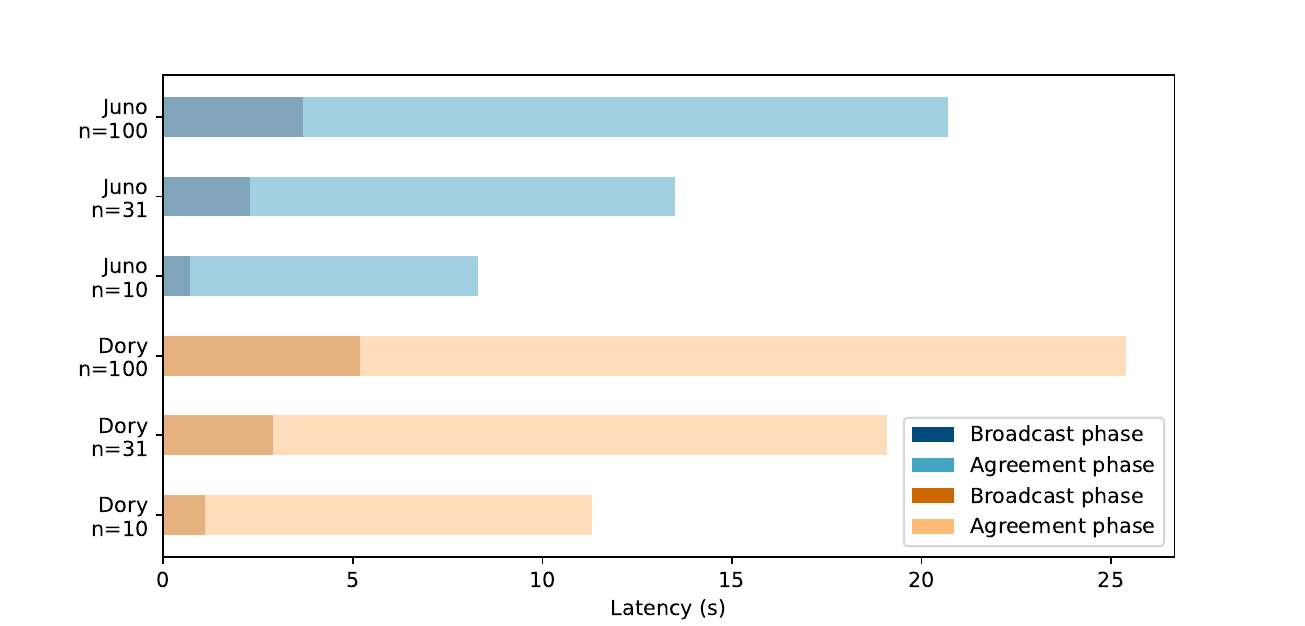}
        \caption{Latency breakdown for \textsc{Juno} \& HoneyBadgerBFT}
        \label{fig:break}
    \end{subfigure}
\caption{Performance evaluation.}
\label{fig:performance_evaluation}
\end{figure}

 \vspace{0.3em}
\noindent\textbf{Throughput analysis} (see Fig.\ref{fig:tput_geo}). Compared to HoneyBadgerBFT, \textsc{Juno} achieves significantly higher throughput rates due to its reduced communication and message complexity. As the batch size increases, the performance disparity becomes even more pronounced. While HoneyBadgerBFT's throughput plateaus, \textsc{Juno}'s throughput continues to scale, demonstrating superior management of larger transaction loads. Particularly in the range of 8,000 to 10,000 transactions, \textsc{Juno} maintains a consistent upward trajectory, indicating its robustness in high-load scenarios. In comparison with Dory, \textsc{Juno} demonstrates superior overall throughput performance. We observe that the performance gap becomes more pronounced as \textit{f} grows. Indeed, a higher number of processes can better leverage the parallel benefits of the vector consensus. In heavy-load scenarios, the performance of \textsc{Juno} slightly drops when the batch size exceeds 7000 in a network with 20 processes. This might occur when the workload surpasses capacity. Additionally, increasing the number of processes may elevate the probability of requiring additional rounds to complete the agreement phase (i.e., exchanging the vote vectors) in \textsc{Juno}. 

\vspace{0.3em}
\noindent\textbf{Latency analysis} (see Fig.\ref{fig:latency}\&Fig.\ref{fig:break}). \textsc{Juno} maintains an overall lower latency than HoneyBadgerBFT (cf. Fig.\ref{fig:latency}), with an improvement margin increasing as the network scales. This improvement is primarily due to the reduced message propagation required to achieve equivalent throughput levels. In comparison with Dory, \textsc{Juno}'s latency is similar when \textit{f} = 1. When the network size is small, the benefits of parallel processing in the vector consensus are not significant. However, as the network size grows, \textsc{Juno} can achieve the same throughput level with a smaller batch size, resulting in approximately 21\% lower overall latency compared to Dory.

To provide a more detailed understanding, we present a latency analysis of \textsc{Juno} and Dory breaking down the two phases of ACS protocols. The batch size is set to 25,000tx (around 6.25MB) to evaluate performance under heavy loads. As depicted in Fig.\ref{fig:break}, latencies in both phases of \textsc{Juno} are lower than those of Dory, especially the broadcast phase. 
Although both protocols use provable broadcast, \textsc{Juno}'s broadcast phase requires far fewer communication rounds. While the agreement phase remains a substantial portion of the overall runtime for both protocols, \textsc{Juno}'s agreement phase is considerably faster due to its message aggregation approach, resulting in fewer communication rounds in normal cases.

%% file: paper/literature.tex

\section{Related Work}
\label{sec-rw}

\noindent\textbf{ACS paradigms}. The asynchronous common subset~\cite{ben1993asynchronous} is a type of Byzantine consensus under asynchronous networks. Each distributed processes propose a value as input and reach an agreement on a common subset (i.e., \textit{n-f}) as output. ACS can be extended to construct many distributed protocols, such as distributed key generation~\cite{kokoris2020asynchronous}\cite{das2022practical}, secret sharing~\cite{das2023practical}, and multi-party computation (MPC)~\cite{chopard2021communication}\cite{lu2019honeybadgermpc}.

In practice, ACS are implemented using two classic paradigms~\cite{zhang2022pace}: BKR and CKPS. The BKR paradigm~\cite{ben1994asynchronous} involves a broadcast instance and an agreement instance for each participant in the network, and has been further refined in subsequent research~\cite{miller2016honey}\cite{duan2018beat}\cite{liu2020epic}. On the other hand, the CKPS paradigm~\cite{cachin2001secure} replaces ABA with MVBA, facilitating consensus on multi-value inputs. This approach has been advanced by a series of works aimed at optimizing performance~\cite{gao2022dumbo}\cite{lu2020dumbo}\cite{guo2022speeding}. Our study  focuses on the BKR paradigm.

\vspace{0.3em}
\noindent\textbf{Asynchronous Byzantine agreements}. 
HoneyBadgerBFT~\cite{miller2016honey} is a recent asynchronous BFT protocol that utilize parallel instances to reduce computing complexity and guarantees liveness without relying on timing assumptions. HoneyBadgerBFT has inspired a series of subsequent works aimed at enhancing its efficiency and scalability. The Dumbo family, for instance, improves upon HoneyBadgerBFT by transitioning from binary to multi-valued Byzantine Agreement (BA)\cite{guo2020dumbo}, incorporating a linear broadcast protocol and a compact consensus protocol\cite{guo2022speeding}. The BEAT family enhances HoneyBadgerBFT through a modular design approach~\cite{duan2018beat}.

Further advancements include PACE~\cite{zhang2022pace}, which introduced a binary BFT protocol with fewer steps per round and a more efficient common coin algorithm. WaterBear~\cite{zhang2023waterbear} extends the security guarantees to partially synchronous settings. FIN~\cite{duan2023fin} offers a signature-free ACS with optimal time complexity.

\vspace{0.3em}
\noindent\textbf{Broadcast protocols.} 
Reliable broadcast protocols are widely used for their message delivery properties in asynchronous settings. RBC protocols exhibit a message complexity of $\mathcal{O}(n^2)$ as a result of the all-to-all communication rounds involved~\cite{bracha1987asynchronous}\cite{aguilera2021frugal}. Provable broadcast protocols, which are employed in pipelined BFT systems~\cite{yin2019hotstuff}\cite{yandamuri2023communication}\cite{decouchant2022damysus}, demonstrate linear message complexity. However, there is an absence of studies examining their potential application within ACS.

\vspace{0.3em}
\noindent\textbf{Vector consensus.} The vector consensus problem, introduced by Doudou et al.~\cite{doudou1998muteness}, is a method for achieving atomic broadcast in distributed systems, where nodes must agree on a vector of values rather than a single value. Neves et al.~\cite{neves2005solving} explored this problem in asynchronous networks, utilizing a ``wormhole'' mechanism to provide reliable and timely hybrid services. Vaidya et al.~\cite{vaidya2013byzantine} demonstrated the impossibility of solving the vector consensus problem without additional support and distinguished between two variants: \textit{exact vector consensus}, applicable in synchronous networks with strict requirements, and \textit{approximate vector consensus}, suitable for asynchronous networks with more relaxed conditions. Additionally, Cachin et al.~\cite{cachin2020anonymity} proposed an anonymity-preserving Byzantine vector consensus protocol by combining ring signatures with vector consensus, thereby reducing the vector consensus problem to a binary consensus problem.